# Imaging Extracellular Protein Concentration with Nanoplasmonic Sensors

Jeff M. Byers[†], Joseph A. Christodoulides[†], James B. Delehanty[†], Deepa Raghu[†], and Marc P. Raphael[†,*]

*Corresponding Author Information:*

Dr. Marc P. Raphael
Code 6363
[†]Naval Research Laboratory
4555 Overlook Ave SW
Washington, DC 20375-5320
PH: 202 767-6096
FAX: 202 404-8933
Email: marc.raphael@nrl.navy.mil

**Abstract**

**Extracellular protein concentrations and gradients queue a wide range of cellular responses, such as cell motility and division. Spatio-temporal quantification of these concentrations as produced by cells has proven challenging. As a result, artificial gradients must be introduced to the cell culture to correlate signal and response. Here we demonstrate a label-free nanoplasmonic imaging technique that can directly map protein concentrations as secreted by single cells in real time and which integrates with standard live-cell microscopes. When used to measure the secretion of antibodies from hybridoma cells, a broad range of time-dependent concentrations was observed: from steady-state secretions of 230 pM near the cell surface to large transients which reached as high as 56 nM over several minutes and then dissipated. The label-free nature of the technique is minimally invasive and we anticipate will enable the quantification of deterministic relationships between secreted protein concentrations and their induced cellular responses.**

From bacterium to eukaryote, a cell's fate is directly tied to its local chemical environment. The measurement of external protein concentrations and gradients by membrane bound receptors has been found to determine the most fundamental of decisions including differentiation (1, 2), motility (3-5) and proliferation (6). For decades such dependencies have been deduced by introducing artificial gradients to cell cultures, yet direct measurements of the spatio-temporal concentrations which cells themselves produce via secretion have remained elusive.

A critical roadblock has been the lack of an assay that can measure extracellular protein concentrations in real time without disrupting the signaling pathways of interest. This real time, non-invasive requirement severely limits the techniques that can be employed, including now commonplace fluorescent labeling methodologies. For instance, while fluorescent fusion proteins have altered the landscape of intracellular protein measurements, the technique does not lend itself to extracellular signaling. It is a considerable challenge to ensure that a 27 kDa tag such as GFP does not compromise the labeled protein's ability to navigate the complexities of the secretory pathway (7, 8). Even if the pairs are successfully secreted, the result is a diffuse fluorescent glow outside the cell which is difficult to quantify. Fluorescently-labeled antibodies used for immunosandwich assays have been successfully introduced outside of live cells to measure secretions (9-11). However, the addition of these relatively large probes (typically 150 kDa) is an impediment to downstream signaling and the techniques typically involve isolating individual cells. In both examples, the ability to establish causal relationships between secreted protein concentrations and cell fate - whether the signaling be autocrine, paracrine or endocrine in nature - is hampered by the probes themselves.

Recent advances in solid-state nanosensors have the potential to overcome this impasse. Probes such as nanodiamonds and metallic nanostructures are biocompatible, do not suffer from photobleaching and, most importantly from the protein secretion perspective, are label-free techniques. Nanodiamonds which incorporate nitrogen vacancies are highly sensitive magnetic field detectors making them particularly applicable to sensing metallo-proteins (12, 13). Metallic nanoparticles exhibit a localized surface plasmon resonance (LSPR) which is sensitive to changes in the local refractive index of the surrounding medium. Their surfaces have been biofunctionalized for the detection of proteins (14-17), lipids (18, 19), and DNA (20, 21) in cell-free environments. In addition, LSPR optical configurations are readily integrated with standard wide-field microscopy setups which has enabled the detection of protein secretions in the presence of hundreds of cells (22, 23) as well as real-time single cell secretions (24). However, measuring extracellular protein concentrations in space and time, a key parameter in the modeling and quantification of signaling pathways, has remained a challenge (25-28).

Here we report the realization of this goal by utilizing arrays of gold plasmonic nanostructures for the real-time imaging of secreted protein concentrations. The inference of concentration from nanoplasmonic imagery was enabled by progress on two fronts. First, we demonstrate that when properly normalized, LSPR imagery (LSPRi) can be used to determine the fraction of active surface ligands bound to the analyte (fractional occupancy). Second, to calculate concentration, we developed an analysis approach based on temporal filtering that utilizes the LSPRi-determined fractional occupancy and reaction rate constants as inputs. We applied this approach to the spatio-temporal mapping of secreted antibody concentrations from hybridoma cells. Single cell secretions were imaged with a time resolution of 15 seconds over a spatial range that extended 130 µm from the center of the cell. Sensing arrays located next to individual cells resolved steady-state concentrations between 0.2 and 1.3 nM. In sharp contrast, burst-like secretions were also measured in which the transient concentrations reached as high as 56 nM over the course of several minutes and then dissipated. We anticipate this ability to measure secreted concentrations with high spatial and temporal resolution will have applicability to numerous analytes and cell types.

**Results and Discussion**

**Calibration of Plasmonic Nanostructures.** Our experiments take place on an inverted wide-field microscope using glass coverslips that have been patterned via electron beam lithography to incorporate arrays of plasmonic gold nanostructures (Fig. 1A). The structures, 75 nm in diameter and 80 nm in height, were arranged in 20 × 20 arrays with a pitch of either 300 nm or 500 nm between nanostructures and 33 µm between arrays, center-to-center. They are illuminated with a 100 W halogen lamp and crossed polarizers are used to minimize background contributions from glass substrate scattered light. In aqueous solutions the arrays have a resonance peak centered at ~ 635 nm. The gold nanostructures are biologically functionalized by first applying a two-component self-assembled monolayer of thiols in a 3:1 ratio. The majority thiol component is terminated with polyethylene glycol to prevent non-specific binding while the slightly longer minority component terminates with an amine group for covalent ligand attachment. Analyte binding to the ligands causes a perturbation in the local index of refraction which is manifested as a spectral red shift and increase in intensity (Fig. 1B inset). When imaged, the arrays are observed to brighten with increasing spectral shift. Our configuration integrates with traditional cell microscopy techniques such as fluorescence and brightfield imaging, which are accessible by the automated switching of a filter cube (24).

In order to infer secreted protein concentration from imagery, the qualitative feature of array brightening on the CCD camera must be quantified in terms of the fractional occupancy, $f$. The law of mass action can then be applied to determine analyte concentration, $C$, using :

$$\dot{f} = k_a C \cdot (1-f) - k_d f \quad\quad\quad [1]$$

where $\dot{f}$ is the time derivative of the fractional occupancy, $k_a$ is the association rate constant and $k_d$ is the dissociation rate constant. To accomplish this calibration we used the setup shown in Fig. 1A in which spectroscopy and imagery are recorded simultaneously for a given array while the analyte is microfluidically introduced. We have previously shown that $f$ can be determined from spectroscopy data by tracking the spectral shift as the analyte concentration is increased from zero to a saturating value (29, 30). Fig. 1B shows an application of this spectrometry-based technique in which fractional occupancy was determined from the introduction of 400 nM of anti-c-myc monoclonal antibodies over c-myc peptide functionalized nanostructures. However, the information gained from binning by wavelength in spectroscopy-based approaches comes at the expense of spatio-temporal resolution. For instance, in the optical configuration of Fig. 1A, the spectral spatio-temporal resolutions were over an order of magnitude lower than those of the CCD camera. To determine the fractional occupancy directly from imagery (LSPRi), the mean array intensity as measured by the camera, $I(t)$, was normalized by $I_N(t) = (I(t) - I_o)/(I_f - I_o)$ where $I_o$ and $I_f$ are the initial and saturated array intensity values (Fig.

1C). When plotted against the spectrally-determined fractional occupancy (Fig. 1D) a linear relationship is evident. This relationship holds whether the analyte is a 150 kDa antibody such as anti-c-myc (red and green data) or 60 kDa neutravidin proteins binding to a biotinylated surface (blue data). If the camera has a strong wavelength-dependence to its quantum efficiency (QE) in the vicinity of the resonance, non-linearities can be introduced. For this reason, we engineered the size and pitch of the nanostructures so that the resonance was located in a relatively flat region of the camera's QE response while also red-shifted from excitation wavelengths used for common fluorescent tags such as GFP and RFP. As a result of this design and calibration, every array in the LSPRi field of view could be used to determine fractional occupancy in real time, without the need for spectrometry.

**Determining Concentration from Fractional Occupancy.** The initial data processing of the LSPR imagery produces an estimated fractional occupancy, $f_i$, and standard deviation, $\sigma_i$, for each of the $M$ images at times $t_i$ (Fig. 1). Eq. 1, however, shows that the concentration is also dependent on the time derivative of the fractional occupancy, $\dot{f}$. A central problem for any data analysis approach that seeks to calculate time-varying concentration is that both $f$ and $\dot{f}$, along with their related uncertainties, must first be jointly determined. The formalism we used to accomplish this can be thought of as divided into three steps. First, a time window, $h$, is defined in which $f$ and $\dot{f}$ are to be calculated, as schematically shown in Fig. 2A. Second, the data within this time window are fit with a set of local linear models dependent on $f$ and $\dot{f}$ (Fig. 2B) and a least-squares approach is used to determine their maximum likelihood values and uncertainties. Finally, the calculated joint probability distribution for $f$ and $\dot{f}$ is combined with Eq. 1 to determine the associated concentration probability distribution for each time window (Fig. 2C).

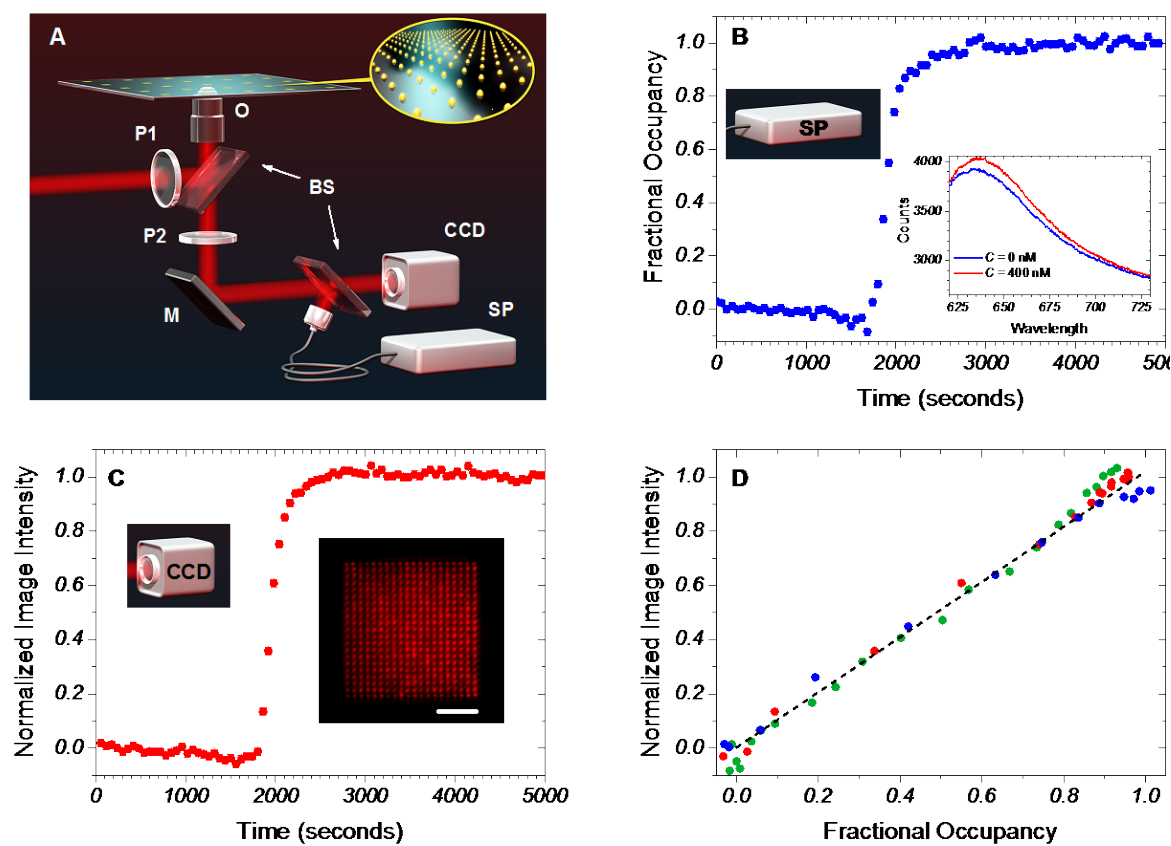

**Fig. 1.** LSPRi calibration using a 400 nanostructure array. (A) Gold nanostructures are patterned atop the coverslip (inset). The excitation light from a halogen lamp passes through a linear polarizer P1 and illuminates the arrays through the objective O. The reflected light is collected by the objective, passed through a crossed linear polarizer (P2) and is reflected by a mirror (M) through a 50/50 beam splitter (BS) to the spectrometer (SP) and CCD camera. (B) A single array is aligned with the optical fiber and the spectra is analyzed to determine the time-dependent fractional occupancy. The inset shows two spectra at concentrations $C = 0$ and $C = 400$ nM. (C) Normalized imagery data on the same array taken in parallel with the spectral acquisition. The inset shows a false colored CCD image of a 20 × 20 array of nanostructures with a pitch of 500 nm (scale bar is 3 µm) (D) Normalized image intensity versus the spectrally-determined fractional occupancy for three separate experiments. The red and green circles are for anti-c-myc monoclonal antibodies binding to a c-myc functionalized array in PBS and serum-free media, respectively. The blue circles are for neutravidin binding to biotinylated nanostructures. The size of the symbols in all plots incorporate $2\sigma$ uncertainty.

In describing the details of this approach some changes in nomenclature are helpful. First, substitute $f_i \rightarrow \mu_i$ to emphasize the connection with the mean parameter of the normal distribution. The processed LSPRi data will then be indicated by $D = \{t_i, \mu_i, \sigma_i \mid i=1, \ldots, M\}$. Also, define a dimensionless concentration: $c = C/K_D$. To summarize the procedure consider the following expression for the probability distribution of the concentration at time, *t*, conditioned on the LSPRi data, $D$, and time window, $h$:

$$p(c|t,h;D) = \frac{1}{Z} \int_0^1 df \int_{-\infty}^{\infty} d\dot{f}\ p(c|f,\dot{f})p(f,\dot{f}|t,h;D) \quad [2]$$

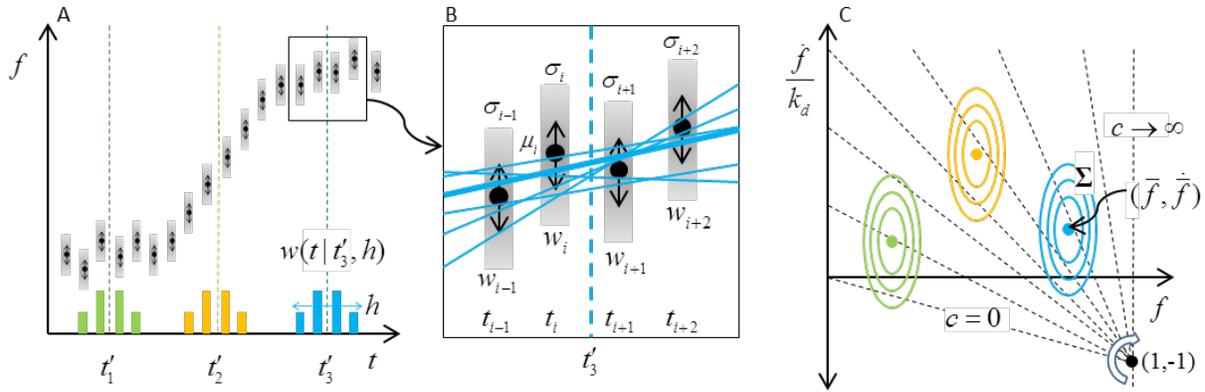

**Fig. 2.** Schematic of Data Analysis to Determine Concentration from Fractional Occupancy. Three steps are needed to determine the probability of a concentration at each time, $t'$: (1) Subsampling the fractional occupancy, (2) forming the probability over parameters of local linear models and (3) integrating along lines of constant concentration. (A) The first step subsamples the processed LSPRi data, $D$, of mean values, $\mu_i$ (black circles) and standard deviations, $\sigma_i$ (grey bars). A temporal filter centered at time, $t$, and width, $h$, assigns weights, $w_i$ (depicted as vertical bars on the $t$-axis), to control the $i^{th}$ sample's contribution to the local linear models. Three different times $(t'_1, t'_2,$ and $t'_3)$ are shown for determination of concentration. (B) A blow-up of the samples around $t = t'_3$ shows local linear models that might fit the data. Given the normal distribution $(\mu_i, \sigma_i)$ for the fractional occupancy at each $t_i$ one can quantify the probability of different local linear models explaining the data. The weights, $w_i$, subsample the data by increasing the variance of data outside the range of $h$ via Eq. 3. Samples not near $t'_3$ are unable to constrain the linear models and do not contribute. (C) Each local linear model is a point in the $f - \dot{f}$ plane. All possible local linear models are summarized by the probability distribution, $p(f, \dot{f} \mid t, h; D)$, a bivariate normal distribution (depicted as elliptical contours) with five parameters: the mean value $(\bar{f}, \bar{\dot{f}})$ and the entries $(\sigma_{xx}, \sigma_{xy},$ and $\sigma_{yy})$ in the 2 × 2 covariance matrix, $\Sigma$. Using the law of mass action for the kinetic binding model, we can assign a concentration to each point $(f, \dot{f})$. The probability of a particular concentration, $c$, at a time $t$ is determined by integrating $p(f, \dot{f} \mid t, h; D)$ along the lines of constant concentration shown as the dashed lines radiating from the point (1,-1). The constant value for the concentration of each line increases in the clockwise direction and each line integral must be successively evaluated to determine $p(c \mid t, h; D)$ for all $c$.

The parameter *h* determines the amount of data to subsample in *D* near the time, *t*, and the normalization, *Z*, is the integral of Eq. 2 over concentration as shown explicitly in the *SI text*. The calculation of $p(c|t,h;D)$ is dependent upon $p(c|f,\dot{f})$ which is described by the kinetics of the reaction (Eq. 1) and $p(f,\dot{f}|t,h;D)$ which, as discussed above, is a central computational challenge given that $\dot{f}$ is not explicitly measured by LSPRi. To determine $\dot{f}$, at a minimum we need to take a numerical derivative without amplifying the noise in the data. Standard practice in time-series analysis uses a smoothing filter over some range of samples in time, reducing the noise in the derivative. However, a better approach is possible in our experiment since we also have a standard deviation, $\sigma_i$, for each $\mu_i$. This allowed us to pose the question, how well can a straight line, $\mu' = f + \dot{f}\cdot(t'-t)$, explain the noisy data near time *t*? Each local linear model with parameters $(f,\dot{f})$ over a range of data prescribed by the filter located at *t* with width, *h*, can be assigned a likelihood of fitting the data.

Similar to linear regression, we can write our probability as a negative log-likelihood, *L*, but with the weights, $w(t_i|t,h)$, of a temporal filter at each $t_i$ of the data,

$$L = -\ln p\left(f,\dot{f}|t,h;D\right) = \sum_{i=1}^{n} w(t_i|t,h) \cdot \frac{\left[f + \dot{f}\cdot(t_i-t) - \mu_i\right]^2}{2\sigma_i^2} + \text{terms ind. of } f \text{ and } \dot{f} \quad [3]$$

If we used the maximum likelihood estimate of *f* and $\dot{f}$ at each time *t* then this technique is identical to re-weighted least squares. Various functions can be selected for the temporal filter and we use a generic Gaussian profile, schematically shown as bar graphs in Fig. 2A, over the data acquisition times, $w(t_i|t,h) = e^{-(t_i-t)^2/2h^2}$, with two adjustable parameters: the center at time *t* and the width *h*. A schematic drawing of linear fits to the data within a chosen filter time window is shown in Fig. 2B. (A discussion of the bias-variance tradeoff regarding the choice of *h* as well as the acausal nature of this filtering approach can be found in the *SI text*, Figs. S1 and S2).

Eq. 3 can be re-written as a bivariate normal distribution function, $p\left(f,\dot{f}|t,h;D\right)$ in terms of five parameters: the mean value $(\bar{f},\bar{\dot{f}})$ and the entries $(\sigma_{xx},\sigma_{xy},\text{ and }\sigma_{yy})$ in the 2 × 2 covariance matrix, $\Sigma$. The probability distribution at each time point can be depicted as elliptical contours in $f - \dot{f}$ plane as shown in Fig. 2C. When inserted into Eq. 1 the result is an integral that can be numerically evaluated at each time *t* over a range of concentrations to estimate the most probable concentration and its associated error (*SI text*).

**Simulated Measurements.** To highlight the general features of the data analysis methodology described above we have simulated concentration data with varying time dependencies. Fig. 3A shows step-wise simulated $C(t)$ data in which the concentration increases slowly, then rapidly, and finally decreases rapidly. For the analysis we use $k_a = 10^6$ M$^{-1}$s$^{-1}$ and $k_d = 10^{-3}$ s$^{-1}$ which are values typical of antibody-antigen interactions.

The fractional occupancy versus time (Fig. 3B) is determined by numerically integrating the differential equation in Eq. 1 forward in time using the initial condition, $f(0) = 0$. Gaussian noise is added to the calculated $f(t)$ using a standard deviation typical of the experimental data shown in Fig. 1B. In Fig 3C, the local linear models (red lines) are displayed for a Gaussian filter with h = 270 s and the resulting calculated concentration in Fig 3D. Because of the relatively high association rate of the receptor-ligand pair, the slow and rapid concentration increases are faithfully reproduced by the analysis with some curvature at the vertices due to the filter width, h. The decreasing concentration step is reproduced but with a time delay of ~ 250 s due to the relatively long receptor-ligand mean binding time, $1/k_d = 1000$ s, which results in delayed sensitivity to sudden decreases in concentration. Increasing $h$ improves the signal to noise ratio of the calculated concentration at the expense of time resolution (see SI text).

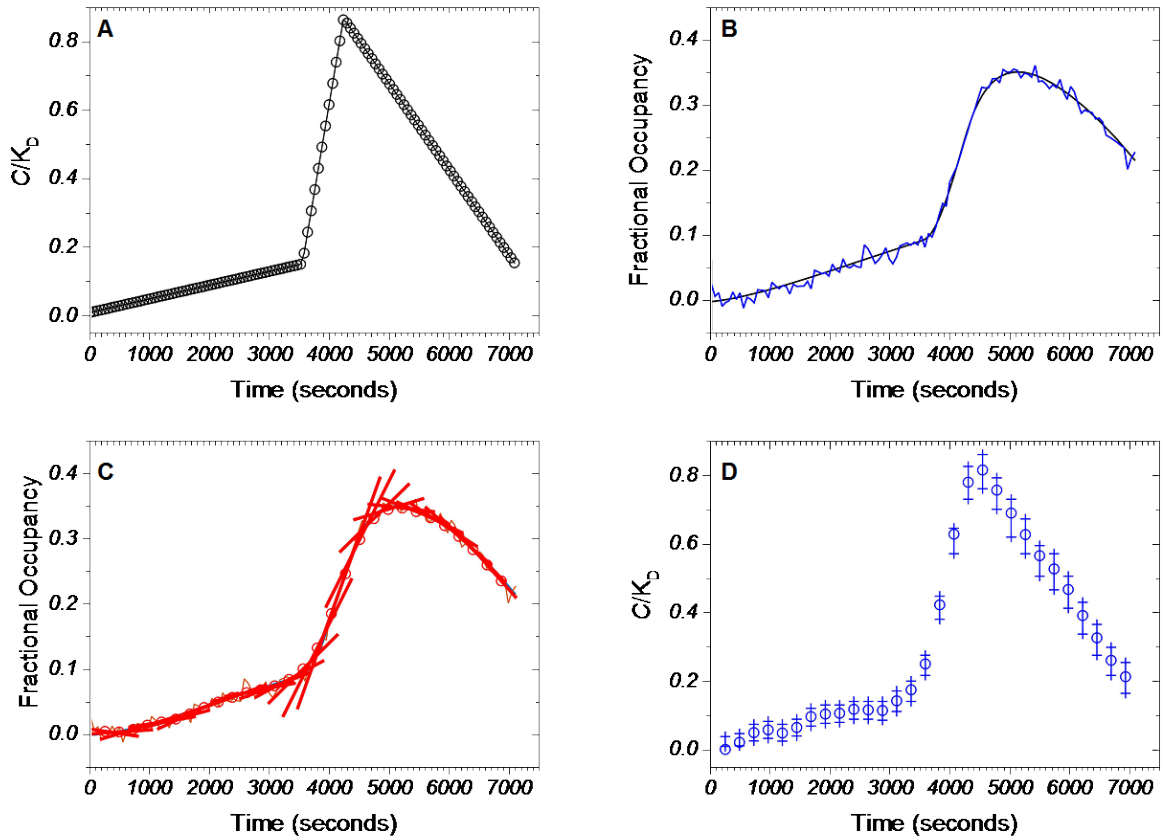

**Fig. 3.** Analysis of simulated concentration data for a receptor-ligand rate constants of $k_a = 10^6$ M$^{-1}$s$^{-1}$, $k_d = 10^{-3}$ s$^{-1}$, $K_D = 1$ nM (A) Piece-wise function of three simulated time-dependent concentration scenarios (B) Time-dependent fractional occupancy as determined by solving Eq. 1 with added Gaussian noise typical of the experimental setup. (C) Local linear model fits to the fractional occupancy for filter width h = 270 s. (D) Calculated concentration. The symbols and error bars represent the calculated mode of the concentration distribution divided by $K_D$ over a 5% to 95% confidence interval.

**Live Cell Secretion Measurements.** Anti-c-myc secreting hybridoma cells were introduced on to a chip with c-myc functionalized nanostructures. The density of cells was adjusted so that the field of view included 2 to 3 cells. At a distance of 70 µm or more from the cells, the secreted antibody concentration fell below the array detection limit (~ 100 pM) allowing for those arrays to be used as negative controls. By having controls in the same field of view, global intensity variations such as those due to focus drift could be subtracted out from the signal of arrays adjacent to cells. At the end of each experiment, a saturating solution of commercial anti-c-myc antibodies was introduced in order to normalize the LSPRi intensity and calculate fractional occupancy. The kinetic rate constants used in the analysis were determined with a commercial SPR instrument using an identical surface functionalization protocol to that of the nanoplasmonic substrates (29): $k_a = 2.68 \times 10^4$ M$^{-1}$s$^{-1}$, $k_d = 4.75 \times 10^{-5}$ s$^{-1}$ and $K_D = k_d/k_a = $ 1.77 nM.

The simultaneous secretion measurements from two cells are shown in Fig. 4. Arrays adjacent to the cells used for the analysis are marked with red and blue boxes; the white box outlines the control array. The time dependent fractional occupancy (Fig. 4B) indicates that the lower cell was secreting at a higher rate than the upper cell. Concentration was determined using a temporal filter with h = 270 s shows a constant concentration over 40 minutes, as expected for a steady state secretion scenario, with an average concentration of 1.30 nM near the lower cell versus 230 pM for the upper cell.

In contrast, the collection of three cells shown in Fig. 5A displayed strongly time dependent secretions. The array to the left of the cells (green outline) measured a rise in fractional occupancy (Fig. 5B) that rose to 0.28 over the course of 2 minutes. This is in sharp contrast from the cells of Fig. 4 in which it took 40 minutes to reach a maximum fractional occupancy of 0.08. The concentration for the green-outlined array, located 24 µm from the center of the three cells, peaked at 56 nM within 2 minutes (Fig. 5C). The rapid increase and decrease in concentration was best resolved using a temporal filter with h = 45 s. The burst was also recorded by the red-outlined array located 43 µm from the center of the three cells. The peak concentration at this array was 9 nM and time-delayed by 91 s from the green-outlined array peak, consistent with a burst of secreted antibodies diffusing outwardly from the three cells. It was unclear from the imagery how many of the cells contributed to the burst.

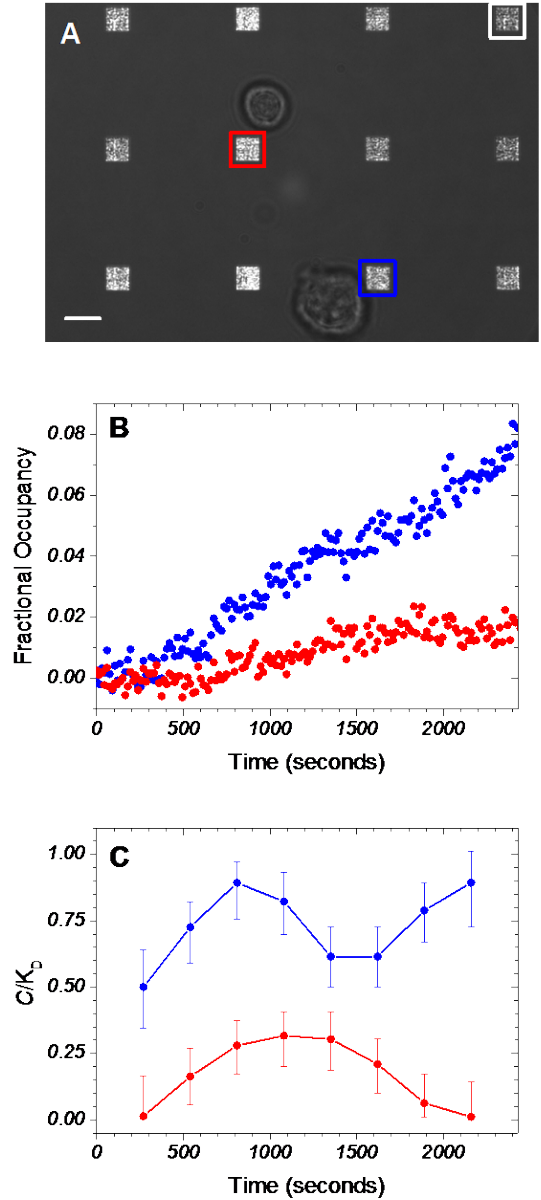

**Fig. 4.** Steady state secretions quantified. (A) Merged LSPRi and brightfield images showing two hybridoma cells amongst 12 arrays. The arrays outlined in red and blue were used to measure the antibody concentration near the upper and lower cells, respectively, while the array outlined in white was used as a control. (B) LSPRi-determined fractional occupancy. Red and blue data points correspond to red and blue outlined arrays from (A) after subtracting control array data. (C) Calculated concentration for the red and blue outlined arrays applying a temporal filter with h = 270 s. The symbols and error bars represent the mode of the concentration probability distribution divided by $K_D$ with a 5% to 95% confidence interval. Scale bar is 8 µm.

A comparison of the $h$ values used in Fig. 4C and Fig. 5C highlights the importance of taking an adaptive approach to the data analysis. The Fig. 4 data, being steady state in nature, can accommodate the longer $h$ value (270 s) without loss of temporal information and take advantage of the improved signal-to-noise (*SI text*). In Fig. 5C, the signal-to-noise is reduced by the shorter $h$ value but the peaks in time are readily resolved. The dynamic range of the sensors is also highlighted by these two figures in which the 56 nM peak of Fig. 5 is 244-fold greater than the concentration measured at the lower cell of Fig. 4A. In general, the optimally designed sensor will have a $K_D$ value centered within the range of possible secretions. Finally, the fact that multiple arrays at varying distances from the cell in Fig. 5 could be utilized to measure the burst secretion underscores the spatial and temporal capabilities of our approach.

**Conclusions**

The results presented here demonstrate the ability of nanoplasmonic imaging to spatially and temporally map the secreted protein concentration from single cells or small groups of cells. We anticipate numerous applications of this technology. In a co-culture environment the label free nature of this measurement enables absolute concentration and concentration gradient measurements from one cell type to be correlated to the response of the other, critical for determining causal relations between the secretions and cellular responses such as motility and division. At the individual cell level, the technique can be used to identify polarized secretions important in developmental biology and cell migration. In addition, the fact that the technique integrates with commonly used techniques in fluorescence microscopy allows for both label and label-free investigations of the cells. Printing applications such as ink jet and dip-pen lithography can be utilized to expand the functionality for multiplexing applications capable of quantifying a variety of secreted proteins in parallel.

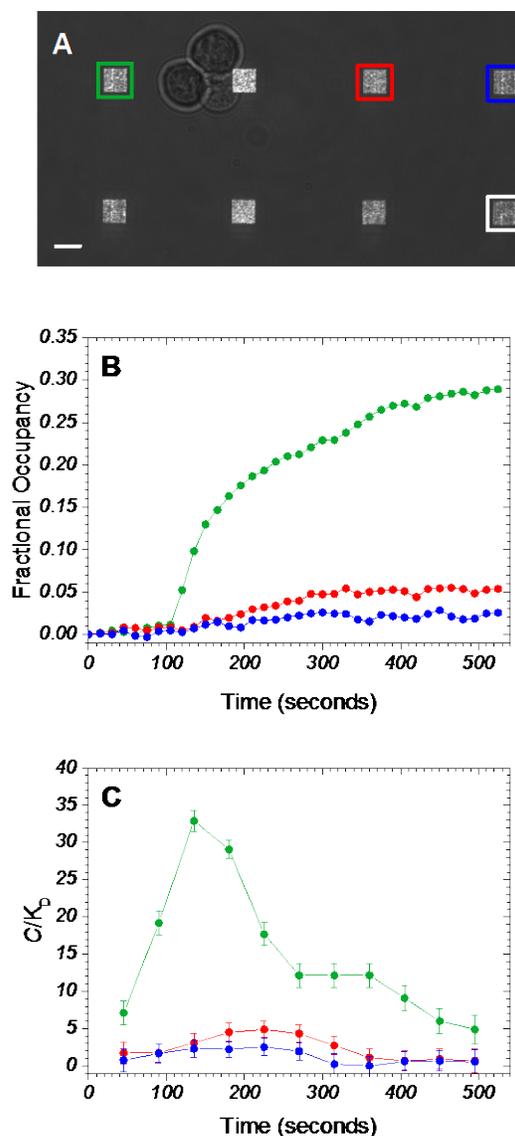

**Fig. 5.** Burst secretion quantified. (A) Merged LSPRi and brightfield images showing a cluster of three hybridoma cells amongst 8 arrays. The arrays outlined in green, red and blue were used to measure the concentration at varying distances from the cell while the array outlined in white was used as a control. (B) LSPRi-determined fractional occupancy. Green, red and blue data points correspond to the green, red and blue outlined arrays in (A) after subtracting the control array data. (C) Calculated concentration for the green, red and blue outlined arrays applying temporal filter analysis with h = 45 s. The symbols and error bars represent the calculated mode of the concentration probability distribution divided by $K_D$ at each time point with a 5% to 95% confidence interval. The scale bar is 8 μm.

## Materials and Methods

**Fabrication of Nanostructure Arrays.** Arrays were patterned onto No. 1.5 glass coverslips by spinning a bilayer resist structure consisting of polymethyl methacrylate and ethyl lactate methyl methacrylate copolymer with thicknesses of 180 nm and 250 nm, respectively. The resist was electron-beam patterned using doses 300 $\mu C/cm^2$ and subsequently developed for one minute in a 2:1 solution of isopropyl alcohol: methyl isobutyl ketone. A 5 nm layer of Ti followed by 70 nm of Au was deposited with a Temescal electron-beam evaporator. The bilayer resist was then lifted off by soaking in acetone for 4 hours.

**Nanostructure Functionalization.** RF plasma ashing (40 W) with 300 mTorr of a 5% hydrogen, 95% argon mixture was used to clean the glass and gold surfaces on the chips. The gold nanostructures were functionalized in a two-component ethanolic-based thiol bath (0.5 mM), consisting of a 3:1 ratio of SH-$(CH_2)_8$-$EG_3$-OH to SH-$(CH_2)_{11}$-$EG_3$-$NH_2$ for 18 hours, where EG stands for ethylene glycol monomer. The amine terminus was reacted with a 10 mg/mL solution of the heterobifunctional crosslinker sulfo-N-succinimidyl-4-formylbenzamide (Solulink) in PBS buffer at pH 7.4, followed by a hydrazine functionalized c-myc peptide conjugation (Solulink) in PBS buffer at pH 6.0 according to the manufacturer's instructions. For biotin-neutravidin studies, 0.3 mM of sulfo-NHS-biotin (Thermo) in PBS was drop coated on to the chip for 30 min. Chips were rinsed with DDW and dried with nitrogen gas. Commercially available monoclonal anti-c-myc antibodies (Sigma) were used for normalizing array response at the end of each experiment.

**Data Analysis.** All analysis was conducted using the Matlab 2013b environment with Curve Fitting, Image Processing and Statistics Toolboxes. Detailed derivations of the equations in the main text and their implementation in Matlab are described in the *SI text*.

**Microscopy Setup and Drift Correction.** Halogen lamp light was first passed through a 594 long-pass filter and then the Koehler illumination train of an inverted microscope (Zeiss AxioObserver) before following the light path described in Fig. 1A. The objective used was a 63X, 1.46 numerical aperture oil-immersion objective. For spectral measurements a 600 µm diameter optical fiber was used to collect the scattered light from a single array and detected with thermoelectrically-cooled, CCD-based spectrophotometer (Ocean Optics QE65000) at an integration time of 1 s. A thermoelectrically-cooled CCD camera (Hamamatsu ORCA R2) with integration times between 200 and 250 ms was used for imagery. A heated stage and temperature controlled enclosure kept the stage temperature at 37.0 ± 0.04 °C (Zeiss). Humidity and $CO_2$ were regulated at 98% and 5%, respectively, by flowing a gas-air mixture though a heated water bottle and into the enclosure. In plane drift was corrected for with image alignment software (Zeiss Axiovision) while the focus was stabilized using an integrated hardware focus correction device (Zeiss Definite Focus).

**Hybridoma Culturing.** Clone 9E10 Hybridoma cells (ATCC) were cultured in complete growth medium RPMI-1640 supplemented with 10% fetal bovine serum and 1% antibiotic/antimycotic in a humidified tissue culture incubator at 37 °C under 5% CO2 atmosphere. Cells were maintained at a density of 3-5 × $10^5$ cells/mL by performing passaging every two days which maintained viability at 90-95%. Prior to LSPRi studies, the cells were pelleted by centrifugation (900 rcf × 5 min) and washed twice with RPMI-1640 SFM for the removal of secreted antibodies and serum. For imaging, 75 µL of 0.5-2 × $10^6$ cells/mL cell solution was manually injected into the fluidics chamber. Cell surface density was controlled by allowing cells to settle on the surface for 5 to 10 min and then microfluidically flowing SFM to remove those still in solution.

**ACKNOWLEDGEMENTS.** We thank George Anderson for helpful comments and discussions. This work was supported by the Naval Research Laboratory's Institute for Nanoscience and a National Research Council Research Associateship Award.

# Imaging Extracellular Protein Concentration with Nanoplasmonic Sensors

Jeff M. Byers, Joseph A. Christodoulides, James B. Delehanty, Deepa Raghu, and Marc P. Raphael

**Determining concentration from fractional occupancy.** The initial data processing of the LSPR imagery (LSPRi) produces an estimated fractional occupancy, $f_i$, and standard deviation, $\sigma_i$, for each of the $M$ images at times $t_i$. The subsequent stages of the data analysis are more complicated and some changes in nomenclature are helpful. First, substitute $f_i \rightarrow \mu_i$ to emphasize the connection with the mean parameter of the normal distribution and prevent confusion with a new model parameter, $f$, without a subscript. The processed LSPRi data will be indicated by $D = \{t_i, \mu_i, \sigma_i \mid i = 1, \ldots, M\}$. Also, define a dimensionless concentration: $c = C/K_D$. To summarize the procedure consider the following expression for the probability distribution of the concentration at time, $t$:

$$p(c \mid t, h; D) = \frac{1}{Z} \int_0^1 df \int_{-\infty}^{\infty} d\dot{f}\ p(c \mid f, \dot{f}) p(f, \dot{f} \mid t, h; D) \quad [\text{S.1}]$$

The parameter $h$ determines the amount of data to subsample in $D$ near the time, $t$, and the normalization, $Z$, is simply the integral of Eq. S.1 over $c$,

$$Z(t, h; D) \equiv \int_0^{\infty} dc\ p(c \mid t, h; D)$$

The procedure is essentially error propagation of the uncertainty in $\mu_i$ as represented by $\sigma_i$ via marginalization (*i.e.*, integrating over the model parameters $f$ and $\dot{f}$, described below) to determine the probability distribution of the concentration, $c$, at each time, $t$, of interest, assuming a particular kinetic binding model represented by $p(c \mid f, \dot{f})$.

The key challenge is to formulate the term, $p(f, \dot{f} \mid t, h; D)$ in Eq. S.1. Kinetic binding models have a time derivative of the fractional occupancy, $\dot{f}$, but the LSPRi image analysis has only provided noisy values of the fractional occupancy, $\mu_i$, at discrete times, $t_i$. At a minimum, we need to take a numerical derivative without amplifying the noise in the data. Standard practice in time-series analysis uses a smoothing filter over some a range of samples in time reducing the noise in the derivative. However, a better approach is possible in our experiment since we also have a standard deviation, $\sigma_i$, for each $\mu_i$. This allowed us to pose the question, how well can a straight line, $\mu' = f + \dot{f} \cdot (t' - t)$, explain the noisy data near time $t$? Each local linear model with parameters $(f, \dot{f})$ over a range of data prescribed by the filter located at $t$ with width, $h$, can be assigned a likelihood of fitting the data. Similar to linear regression, we can write our probability as a negative log-likelihood, $L$, but with the weights, $w(t_i \mid t, h)$, of a temporal filter at each $t_i$ of the data,

$$L = -\ln p\left(f, \dot{f} \mid t, h; D\right) = \sum_{i=1}^{n} w(t_i \mid t, h) \cdot \frac{\left[f + \dot{f} \cdot (t_i - t) - \mu_i\right]^2}{2\sigma_i^2} + \text{terms ind. of } f \text{ and } \dot{f} \quad [\text{S.2}]$$

Notice that the weights are effectively changing the variance of each sample in the local linear model. As $w_i \to 0$ for a sample that sample acquires a large variance and does not constrain the choice of parameters for the linear models formed at $t$. If we used the maximum likelihood estimate of $f$ and $\dot{f}$ at each time $t$ then this technique is identical to re-weighted least squares. However, this point estimate only has value if we can also quantify its uncertainty and this necessitates a more detailed probabilistic model. Various functions can be selected for the temporal filter and we use a generic Gaussian profile over the data acquisition times, $w(t_i | t, h) = e^{-(t_i-t)^2/2h^2}$, with two adjustable parameters: the center at time $t$ and the width $h$. A different symmetric, location-scale function (e.g., Lorentzian, Epanechnikov) can be chosen as the filter with little change in the results. The only constraints are that the function needs to be positive and have a maximum value of one.

The width, $h$, is a free parameter that can be fixed for the entire data set or adaptively for each $t$ in a more sophisticated algorithm. We have opted for selecting a single value of $h$ with an eye towards balancing the effects of small and large values of $h$. The statistical property of bias-variance tradeoff is the key consideration. A narrow width (small $h$) of sampling in time provides a very local estimate of $f$ and $\dot{f}$ but high variance due to the small number of noisy samples. A wider width (large $h$) samples more data and reduces the variance but the bias will increase if non-linearites in $f$ and $\dot{f}$ emerge on larger time scales. In practice, setting $h$ is not difficult and we have used synthetic data to gain a better understanding of this process (see the next section of the Supplementary material).

A technical note, since samples are used both backward ($t_i \leq t$) and forward ($t_i > t$) in time, this is *not* a causal filter and can only be used after all the data has been taken. Obviously, a causal filter can only use data in the past, requiring the filter to have weights $w(t_i > t) = 0$. This would increase the uncertainty in determining the concentration with no advantage in our experiments since we are not attempting real-time control of the environment using the measurements.

We use Laplace's method (a Taylor series expansion of $L$ to 2nd order at the maximum value of $L$) to rewrite the probability distribution as a bivariate normal distribution with five parameters $(\bar{f}, \bar{\dot{f}}, \rho_{xx}, \rho_{xy}, \rho_{yy})$:

$$L = \text{const.} + \frac{1}{2}\begin{pmatrix} f-\bar{f} \\ \dot{f}-\bar{\dot{f}} \end{pmatrix}^T \Sigma^{-1} \begin{pmatrix} f-\bar{f} \\ \dot{f}-\bar{\dot{f}} \end{pmatrix}, \quad \Sigma^{-1} = \begin{pmatrix} \sigma_{xx} & \sigma_{xy} \\ \sigma_{xy} & \sigma_{yy} \end{pmatrix}^{-1} = \begin{pmatrix} \rho_{xx} & \rho_{xy} \\ \rho_{xy} & \rho_{yy} \end{pmatrix} \quad [\text{S.3}]$$

Taking the first derivatives of $L$ (in Eq.S.2) with respect to $f$ and $\dot{f}$, and setting these to zero yields a set of equations for the location, $\bar{f}$ and $\bar{\dot{f}}$, of the maximum value of $L$:

$$\left.\frac{\partial L}{\partial f}\right|_{f=\bar{f}, \dot{f}=\bar{\dot{f}}} = \sum_{i=1}^{n} w_i \cdot \frac{\bar{f}+\bar{\dot{f}}\cdot(t_i-t)-\mu_i}{\sigma_i^2} = 0$$

$$\left.\frac{\partial L}{\partial \dot{f}}\right|_{f=\bar{f}, \dot{f}=\bar{\dot{f}}} = \sum_{i=1}^{n} w_i \cdot (t_i-t) \cdot \frac{\bar{f}+\bar{\dot{f}}\cdot(t_i-t)-\mu_i}{\sigma_i^2} = 0 \quad [\text{S.4}]$$

Taking the second derivatives of $L$ provides the equations for the inverse covariance matrix,

$$\rho_{xx} = \frac{\partial^2 L}{\partial f^2} = \sum_{i=1}^{M} \frac{w_i}{\sigma_i^2}, \quad \rho_{xy} = \frac{\partial^2 L}{\partial f \partial \dot{f}} = \sum_{i=1}^{M} \frac{w_i}{\sigma_i^2} \cdot (t_i - t), \quad \text{and} \quad \rho_{yy} = \frac{\partial^2 L}{\partial \dot{f}^2} = \sum_{i=1}^{M} \frac{w_i}{\sigma_i^2} \cdot (t_i - t)^2$$

There are no further terms that depend on $f$ and $\dot{f}$, and so our parameterization as a bivariate normal distribution is exact for linear models.

The second derivatives can be used to re-write Eq. S.4 as

$$\rho_{xx}\bar{f} + \rho_{xy}\bar{\dot{f}} = \underbrace{\sum_{i=1}^{M} \frac{w_i}{\sigma_i^2} \mu_i}_{\equiv a} \qquad \rho_{xy}\bar{f} + \rho_{yy}\bar{\dot{f}} = \underbrace{\sum_{i=1}^{M} \frac{w_i}{\sigma_i^2} \mu_i \cdot (t_i - t)}_{\equiv b}$$

and solved to obtain

$$\bar{f} = \frac{\rho_{yy} a - \rho_{xy} b}{\rho_{xx}\rho_{yy} - \rho_{xy}^2} \quad \text{and} \quad \bar{\dot{f}} = \frac{\rho_{xx} b - \rho_{xy} a}{\rho_{xx}\rho_{yy} - \rho_{xy}^2}$$

We now have a bivariate normal distribution for $p(f, \dot{f} \mid t, h; D)$ that can be evaluated in the integrals of Eq. S.1,

$$p(f, \dot{f} \mid t, h; D) \cong \frac{(\rho_{xx}\rho_{yy} - \rho_{xy}^2)^{1/2}}{2\pi} \cdot \exp\left[-\frac{1}{2}\begin{pmatrix} f - \bar{f} \\ \dot{f} - \bar{\dot{f}} \end{pmatrix}^T \begin{pmatrix} \rho_{xx} & \rho_{xy} \\ \rho_{xy} & \rho_{yy} \end{pmatrix} \begin{pmatrix} f - \bar{f} \\ \dot{f} - \bar{\dot{f}} \end{pmatrix}\right]$$

All the parameters, $(\bar{f}, \bar{\dot{f}}, \rho_{xx}, \rho_{xy}, \rho_{yy})$, are expressed in terms of the weights at time $t$, $w(t_i \mid t, h)$, and the processed LSPRi data, $D = \{t_i, \mu_i, \sigma_i \mid i = 1, \ldots, M\}$. Note that the local linear model can be constructed around any value of $t$, not just at those times when data was acquired (as in the initial LSPRi analysis).

The probability $p(c \mid f, \dot{f})$ represents the relationship of the fractional occupancy to the concentration and is, therefore, the contribution from the kinetic binding model. Since we are using the Law of Mass Action ($\dot{f}(c, f) = k_d c - k_d(1 + c) \cdot f$) there is a deterministic equation that relates these quantities:

$$c = \gamma(f, \dot{f}), \quad \text{where} \quad \gamma(f, \dot{f}) \equiv \frac{k_d^{-1}\dot{f} - f}{1 - f}$$

This means that we can write $p(c \mid f, \dot{f}) \propto \delta(c - \gamma(f, \dot{f}))$ and reduce our two-dimensional integral to a line integral:

$$p(c \mid t, h; D) = \frac{1}{Z} \int_0^1 df \int_{-\infty}^{\infty} d\dot{f}\, \delta(c - \gamma(f, \dot{f})) p(f, \dot{f} \mid t, h; D)$$

$$= \frac{1}{Z}\sqrt{1 + (1 + c)^2} \int_0^1 df\, p(f, \dot{f}(c, f) \mid t, h; D)$$

where $\dot{f}(c,f) = k_d c - k_d(1+c) \cdot f$. In Fig. S1 the geometry of the integration is shown. The pre-factor $\sqrt{1+(1+c)^2}$ results from the change of variables to form the differential line element as a function of $f$ only.

The final step is to evaluate the integral numerically at each time $t$ over enough values of $c$ to estimate the width of the probability distribution and, thus, the associated error. By completing the square and some algebraic manipulations, the integral can be expressed as:

$$p(c\,|\,t,h;D) = \frac{1}{Z(t,h;D)}\sqrt{1+(1+c)^2}\,e^{-\frac{1}{2}C+\frac{B^2}{2A}}\int_0^1 df\; e^{-\frac{A}{2}\left(f-\frac{B}{A}\right)^2}$$

where the coefficients, $A$, $B$, and $C$, are functions of the concentration and the parameters of the bivariate normal distribution $(\bar{f},\dot{\bar{f}},\rho_{xx},\rho_{xy},\rho_{yy})$ but independent of $f$:

$$A(c) = \rho_{xx} - 2k_d\rho_{xy}(1+c) + k_d^2\rho_{yy}(1+c)^2$$

$$B(c) = \rho_{xx}\bar{f} + k_d\rho_{xy}\left(k_d^{-1}\dot{\bar{f}} - c - \bar{f}\cdot(1+c)\right) - k_d^2\rho_{yy}\left(k_d^{-1}\dot{\bar{f}} - c\right)\cdot(1+c)$$

$$C(c) = \rho_{xx}\bar{f}^2 + 2k_d\rho_{xy}\bar{f}\left(k_d^{-1}\dot{\bar{f}} - c\right) + k_d^2\rho_{yy}\left(k_d^{-1}\dot{\bar{f}} - c\right)^2$$

These expressions have been written to emphasize that each is dimensionless. Note that he term $C$ is not related to the concentration in the main body of the paper.

The Gaussian integral over the interval 0 to 1 can be solved in terms of error functions, erf($x$), however, it is just as easy to evaluate using numerical integration. The integral was solved using the *integral* function in the MATLAB 2013b environment that employs globally adaptive quadrature. This is repeated on an evenly-spaced logarithmic (log$_{10}$) grid of 500 values of $c$ ranging from $10^{-4}$ to $10^5$ for each time $t$. The normalization constant, $Z$, is computed by non-adaptive numerical integration using only these values of $c$. The resulting probability distributions $p(c\,|\,t,h;D)$ can be summed over sub-intervals of $c$ to produce confidence intervals at each time $t$, typically at 5% and 95% of the total probability.

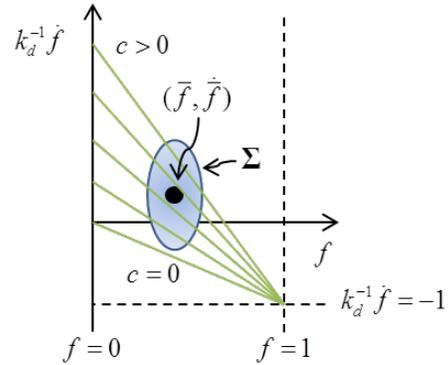

**Fig. S1.** Schematic of the integration domain of $p(c\,|\,t,h;D)$ over $f$ and $\dot{f}$. The ellipsoid represents the probability of different local linear models, $p(f,\dot{f}\,|\,t,h;D)$. The covariance, $\Sigma$, is merely the inverse of the matrix in Eq. S.3 containing the $\rho$'s. Each line represents the path of the line integral as prescribed by the Law of Mass action kinetics for a given value of the concentration, $c$, at time $t$.

**Signal to Noise versus $h$ value.** The analysis technique for inferring concentration is adaptive in the sense that the width of the Gaussian filter, as described by $h$, can be adjusted to best accommodate the data. Longer $h$ values enhance the signal to noise ratio (S/N) at the expense of reducing the temporal resolution. As an example we plot the calculated concentration for the blue-outlined array in Fig. 4 of the main text for h = 270 s and h = 150 s in Fig S2. The concentration remains the same but the error is considerably less for the h = 270 s data points. Because of the steady state nature of the secretion, the error bars overlapped for all the data and no temporal information was lost by using the longer h value. In general the S/N increases linearly with increasing h values. This is shown in Fig S3, again using the data from the blue-outlined array in Fig 4 of the main text, for h values ranging from 120 s to 270 s. The signal was defined as the average of $C/K_D$ values and the noise as average of the associated 5% to 95% confidence intervals.

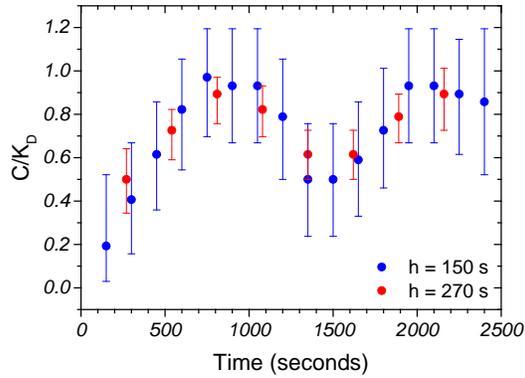

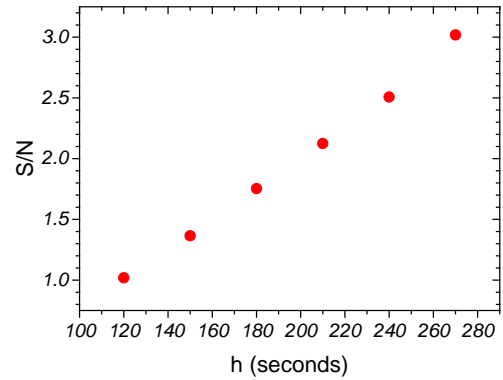

**Fig S2.** Analysis of the data from the blue-outlined array in Fig. 4 of the main text using h = 150 s and h = 270 s.

**Fig. S3.** Signal to Noise (S/N) versus h values for the calculated $C/K_D$ of the blue-outlined array in Fig. 4 of the main text.